\documentclass[amsmath,amssymb,aps,pra,10pt,showpacs]{revtex4-1}

\usepackage{amsmath}

\usepackage[dvips]{color}
\usepackage[normalem]{ulem}
\usepackage{epsfig}\usepackage{epsf}\usepackage[babel=true,kerning=true]{microtype}\usepackage{soul}\usepackage{xcolor}\usepackage{ulem}

\newcommand{\stkout}[1]{\ifmmode\text{\sout{\ensuremath{#1}}}\else\sout{#1}\fi}
\makeatother

\begin{document}

\title{Geometrical optics limit of phonon transport in a channel of disclinations}

\author{S\'{e}bastien Fumeron, Bertrand Berche}

\address{ Statistical Physics Group, IJL, UMR Universit\'{e} de Lorraine - CNRS
7198 BP 70239, 54506 Vand\oe uvre les Nancy, France}

\author{Fernando Moraes}
\address{Departamento de F\'{\i}sica, CCEN, Universidade Federal da Para\'{\i}ba,
Caixa Postal 5008, 58051-900, Jo\~ao Pessoa, PB, Brazil}
\address{Departamento de F\'{\i}sica, Universidade
    Federal Rural de Pernambuco, 
    52171-900 Recife, PE, Brazil}

\author{Fernando A. N. Santos}
\address{Departamento de Matem\'atica  Universidade Federal 
de Pernambuco, 50670-901, Recife, PE, Brazil }

\author{Erms Pereira}

\address{Escola Polit\'ecnica de Pernambuco, Universidade de
Pernambuco, Rua Benfica, 455, Madalena, 50720-001 Recife, PE, Brazil }

 \begin{abstract}
The presence  of topological defects in a material
can modify its electrical, acoustic or thermal properties. However,
when a group of defects is present, the calculations
can become quite cumbersome due to the differential equations that can emerge from the modeling.
In this work, we express phonons as geodesics of a $2+1$ spacetime in the presence of a channel of dislocation dipoles in a crystalline environment described analytically in the continuum limit with differential geometry methods. We show that such a simple model of 1D array of topological defects
  is able to guide phonon waves. The presence of defects indeed distorts the effective metric of the material, leading to an anisotropic landscape of refraction index which curves the path followed by phonons, with focusing/defocusing properties
 depending
on the angle of the incident wave. As a consequence, using Boltzmann transfer equation, we show that the defects may induce an enhancement or a depletion of the elastic energy transport.  We comment on the possibility of designing artificial materials through the presence of topological defects.\end{abstract}



\maketitle

\section{Introduction}

Influence of elastic defects on condensed matter systems \cite{Ran09, Lund15} and the possibility that they offer to design artificial materials \cite{Seebauer10, Carr10, Yoshida15} are still raising a considerable attention. In elastic solids, the feasibility to engineer lattice defects at a molecular scale opened the possibility to tune heat transport at an unprecedented level, as prescribed by the emerging field of phononics \cite{Maldovan13}. For example, in graphene sheets, the presence of defects (such as vacancies, Stone-Wales defects...) was shown to drastically reduce thermal conductivity \cite{Haskins11}, but it could also be used to design thermal junction with rectification ratio up to 46\% \cite{Zhao15}.

\begin{figure}[h!]
\centering \includegraphics[height=1.5in]{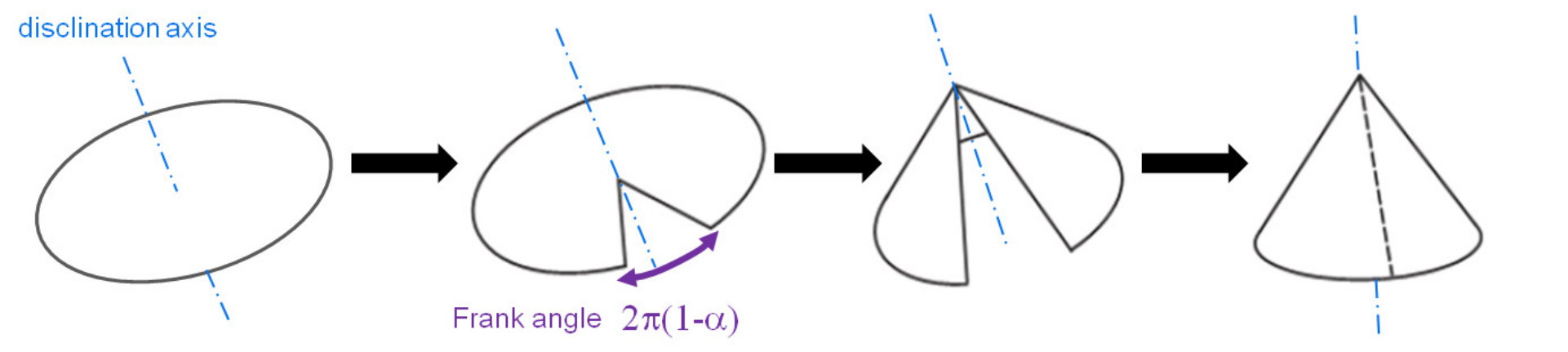} \caption{Volterra cut-and-glue process for a disclination of rank parameter $\alpha<1$. Starting from a flat non-defected medium, removing of a angular wedge of $2\pi(1-\alpha)$ (known as Frank angle) and then gluing the remaining lips together leads to a cone-like geometry.}
\label{discli} 
\end{figure}

Instead of using a complicated set of boundary conditions (as usually done in elasticity theory), pioneering works \cite{Bilby55, Kroner60, Katanaev92} showed that a topological defect is more adequately described as a distorted manifold. As testified by analog gravity models \cite{Moraes00,Barcelo05}, the relevance of differential geometry does not indeed restrict to gravitation, but it can also be used for a wide range of classical and quantum systems such as liquid crystals \cite{Pereira13}, superfluid helium \cite{Volovik03}, 2D electrons gas \cite{Vozmediano,Sinha,ElasticLL}, biological nanostructures \cite{MedinaChiral} and propagation of light in non-homogeneous \cite{FumeronAsllanaj08} or anisotropic \cite{metamats} media, for instance. A disclination is a line-like elastic defect associated with a breaking of rotational symmetry around its axis and the surface density of disclinations identifies with the Riemann curvature tensor \cite{Katanaev05}. The effective background geometry representing the defect can be understood as an initially flat geometry (the non-defected state) from which a wedge of material (of Frank angle $F=2\pi(1-\alpha)$) was either removed or added, leading to either a conical ($\alpha<1$) or an hyperbolic geometry ($\alpha>1$) (see Fig. \ref{discli}). 
In a previous work \cite{Fumeron08}, one of us showed that in the vicinity of a single disclination, phonons (quanta of lattice vibrations) experience the defected medium as an anisotropic inhomogeneous crystal. They no longer propagate along straight lines and they may undergo total internal reflections that prevent them from approaching too close to the defect core (the shortest distance a phonon can approach the core is typically of the order of the lattice parameter, that is about 1 nm). Moreover, the radiative power carried by phonons (or radiance) is governed by a modified Boltzmann Transport Equation (BTE), such that radiance can be locally enhanced by a deficit angle disclination.

Of course, in real three-dimensional solids, one cannot consider heat conduction around a single disclination due to the very high elastic energy cost needed to create a unique defect. However, elastic energy levels can be strongly reduced when two disclinations of opposite Frank angles gather into dipoles, leading to stable self-screened configurations \cite{Romanov09}. Such dipoles were recently shown to proliferate along grain boundaries in copper, recrystallized titanium, inclusion-free steel or olivine-rich layers of the Earth mantle \cite{Cordier14}  For instance, in pure copper, where the lattice spacing is about $0.36$ nm, the dipole separation is of the order of $1\mu$m for a disorientation angle ranging between $55^{\text{o}}$ and $62^{\text{o}}$ \cite{Beausir13}. These circumstances lead to one-dimensional alignments of defects (alleys) which break the possible rotational invariance of the material.

Other realistic systems are graphene and related quasi-two-dimensional materials, such as boron nitride, silicene, germanene, etc.. Here, since the structure can relax by buckling into the third dimension, the defects are more likely to occur. As an example and possible application of the methods described here, we depict in Fig. 2 two rows of alternating positive and negative Frank angle disclinations, respectively, pentagons and heptagons. These defect dipoles are the basis of the Stone-Wales defects in graphene \cite{Ma}. Referring to Fig. 2, we have that $\alpha=5/6$, $F=\pi /3$ for a pentagon, and $\alpha=7/6$, $F=-\pi /3$ for a heptagon disclination, on the graphene lattice. The accepted  $C-C$ bond length for graphene is $\ell$ = 0.142~nm \cite{Delhaes}. This gives an estimation of the scale of the defect array shown in Fig. 2. That is, the approximate distance, $2a$, between the centers of two consecutive heptagons (or two consecutive pentagons) in the same row is $2a \approx 2\sqrt{3}\ell \approx 3.5 \ell$. Using for the width of the street the distance, $2b$, between the line passing trough the centers of the pentagons in one row and the line through the centers of the heptagons in the other row, we have $2b \approx 3.5 \ell$ as well. Since we are not focusing on any specific material, without loss of generality, in what follows, we use a continuum version of a disclination array illustrated in Fig. 3, where we make $a=1$ and $b=0.5$ in arbitrary units. 
\begin{figure}
\includegraphics[scale=0.24]{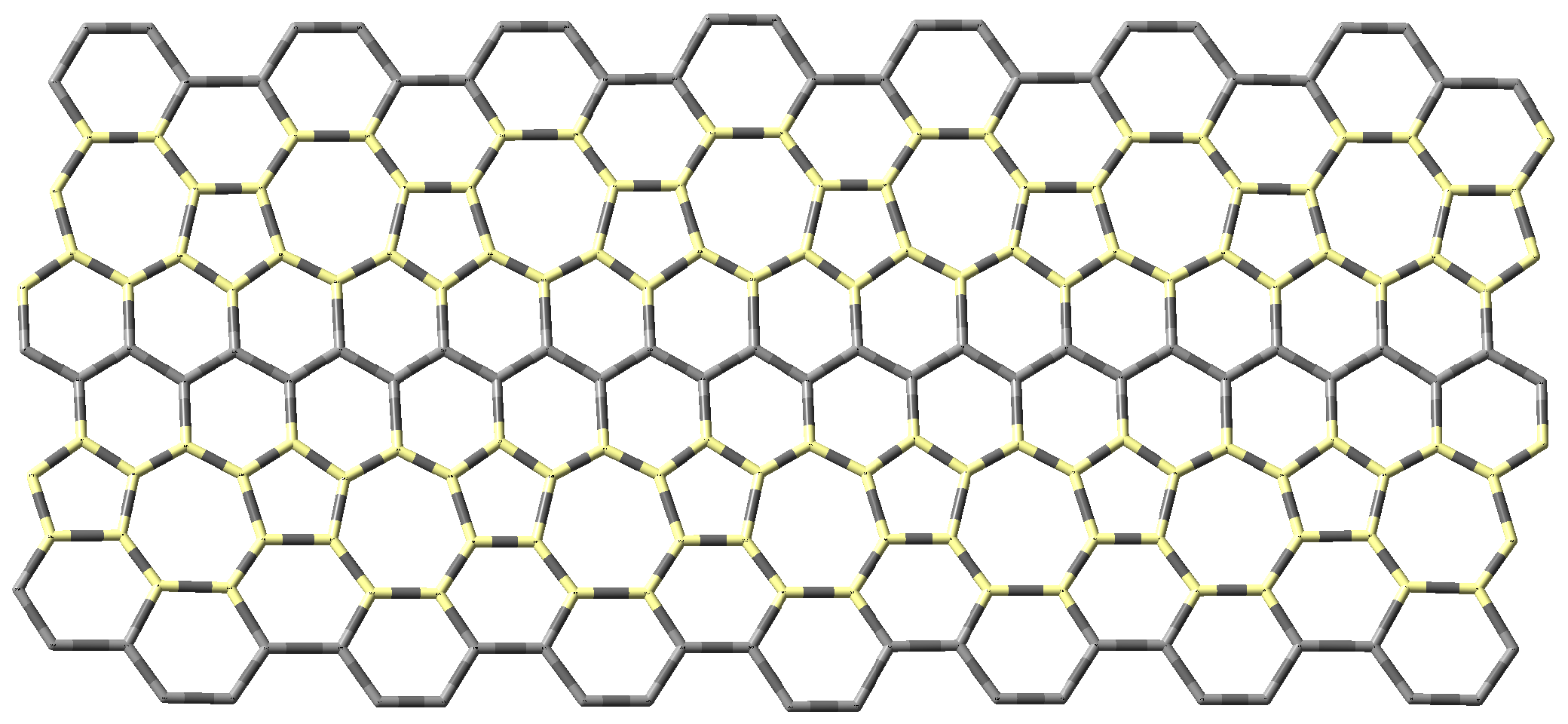} \label{carbonstreet}
\caption{Disclination ``street'' of alternating heptagons and pentagons in an otherwise all-hexagon carbon lattice. All vertexes represent the carbon atom sites. The two defect lines are highlighted for easy identification of the pentagons and hexagons.} 
\end{figure}

In this work, we study the transport of scalar bulk waves in the presence of a channel of disclination dipoles. In the first part, following Letelier  \cite{Letelier01}, we build a geometric model for the distribution of alternate disclinations (dipoles), and the main properties of geodesic paths followed by the waves (in the geometrical optics limit) are discussed. As far as we know, the associated effective metric is calculated here for the first time. Then, we examine the foundations of the radiative transfer equation for the specific intensity from the standpoint of kinetic theory. Finally, the complete form of the stationary BTE in the presence of the defects is presented and analysed, allowing for a discussion of focusing/defocusing of
the energy in the system.

\section{The channel of disclinations geometry}

\begin{figure}[h!]
\centering 
\includegraphics[width=1\linewidth]{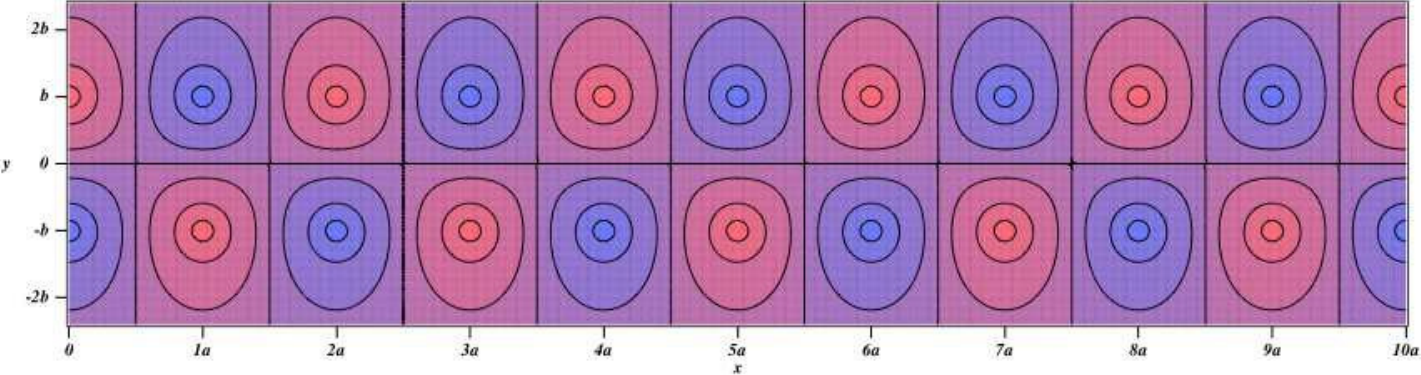} \vspace{-1cm}
\caption{A representation of the distribution of alternate disclinations given by the contour plot of the ``potential" (\ref{fhat}). The positive disclinations correspond to red contours while negative disclinations correspond to blue contours.}  
\label{row}
\end{figure}

From the standpoint of differential geometry, we wish to describe a channel of disclination dipoles, consisting in two infinite rows made of alternate disclinations separated by distance $2a$, the distance between the rows being $2b$. As illustrated in Fig.\ref{row}, the positive disclinations (red contours) are at points $(na,(-1)^{n}b),n\in\mathbb{Z}$, while negative disclinations (blue contours) have coordinates $(na,(-1)^{(n+1)}b),n\in\mathbb{Z}$.
\noindent 
As always done in analog gravity models, the bulk medium is considered in the continuum limit (\textit{i.e.}  limit of a vanishing lattice spacing). Then, adapting to our case the  procedure given in \cite{Letelier01}, the corresponding background geometry is generally described by the spacetime line element
\begin{equation}
ds^{2}=-c^{2}dt^{2}+e^{-4V(x,y)}\left(dx^{2}+dy^{2}\right)+dz^{2}=g_{\mu\nu}dx^{\mu}dx^{\nu}. \label{metric}
\end{equation}
$c$ would be the velocity of light in the general relativity context, but here it stands for the local speed of wave packet, $g_{\mu\nu}$ is the metric tensor (Greek indices run from 0 (time index) to 3 and the Einstein summation convention on repeated indices is used \cite{Schutz}). Since we consider {\it non shifted} disclination lines, the metric will eventually differ from that of Ref. \cite{Letelier01}, hence we present below the whole calculation.  In the function 
\begin{equation}
V(x,y)=(1-\alpha)f_{V}(x,y), \label{V}
\end{equation}
$f_{V}$ describes the space distribution of the defects and $\alpha$ is related to the topological charge of a single disclination. In the case studied here, the topological charge is the Frank angle. In the case studied in Ref. \cite{Letelier01}, the defects are cosmic strings and $(1-\alpha)$ represents the linear mass density of a single string.

In the case of a single row of alternate $\pm F$ defects at $y=b$ (positive disclinations at $..-4a,-2a,0,2a,4a...$ and negative disclinations at $..-3a,-a,a,3a...$), this function writes 
\begin{eqnarray}
f_{V}(x,y)  = \frac{1}{2}\sum_{p=-\infty}^{+\infty} \left\{ \ln\left[\left(x-2pa\right)^{2}+(y-b)^{2}\right]-\ln\left[\left(x-[2p-1]a\right)^{2}+(y-b)^{2}\right]\right\}\qquad \label{eqf} \\
\phantom{f_{V}(x,y) } = \frac{1}{2}\sum_{n=-\infty}^{+\infty} \left(-1\right)^{n}\ln\left[\left(x-na\right)^{2}+(y-b)^{2}\right], \label{sum-glob}
\end{eqnarray}
where the minus sign in front of $\ln$ in Eq. (\ref{eqf}) comes from the negative Frank angle defects. This way, for the alternate row of defects, we have 
\begin{equation}
V(x,y)=\frac{|F|}{2\pi}f_{V}(x,y),  \label{Vnew}
\end{equation}
whith $f_V(x,y)$ given by (\ref{sum-glob}).

Introducing the complex variable $\xi=x+i(y-b)$, one obtains 
\begin{eqnarray}
2f_{V}(x,y) & = & \sum_{n=-\infty}^{+\infty}\left(-1\right)^{n}\ln\left[\left(\xi-na\right)\left(\bar{\xi}-na\right)\right]\label{fv1}\\
 & = & \ln\left(\xi\bar{\xi}\right)+\sum_{n=1}^{+\infty}\left(-1\right)^{n}\ln\left[\left(\xi^{2}-n^{2}a^{2}\right)\left(\bar{\xi}^{2}-n^{2}a^{2}\right)\right]\\
 & = & 2\hat{f}_{V}+4\sum_{n=1}^{+\infty}\ln(an),\label{fv11}
\end{eqnarray}
where $\bar\xi$ denotes the complex conjugate and the function $\hat{f}_{V}$ can be obtained by separating odd
and even contributions in (\ref{sum-glob}) 
\begin{eqnarray}
2\hat{f}_{V}(\xi) & = & \ln\left[\xi\prod_{m=1}^{+\infty}\left(1-\frac{\xi^{2}}{4m^{2}a^{2}}\right)\bar{\xi}\prod_{p=1}^{+\infty}\left(1-\frac{\bar{\xi}^{2}}{4p^{2}a^{2}}\right)\right]\nonumber \\
 &  & -\ln\left[\prod_{m=1}^{+\infty}\left(1-\frac{\xi^{2}}{(2m-1)^{2}a^{2}}\right)\prod_{p=1}^{+\infty}\left(1-\frac{\bar{\xi}^{2}}{(2p-1)^{2}a^{2}}\right)\right].\label{fv2}
\end{eqnarray}
It appears \cite{Letelier01} that the singular structure is
left invariant when adding a constant to $\hat{f}_{V}$. Therefore
(\ref{fv11}) and (\ref{fv2}) represent the same geometry up to a
scaling factor. By using the identities 
\begin{equation}
\sin(x)=x\prod_{m=1}^{+\infty}\left[1-\left(\frac{x}{m\pi}\right)^{2}\right],\;\;\;\cos(x)=\prod_{m=1}^{+\infty}\left[1-\left(\frac{2x}{(2m-1)\pi}\right)^{2}\right],
\end{equation}
it finally comes that 
\begin{eqnarray}
\hat{f}_{V}(x,y) & = & \ln\left|\sin\left(\frac{\pi}{2a}\left[x+i(y-b)\right]\right)\right|-\ln\left|\cos\left(\frac{\pi}{2a}\left[x+i(y-b)\right]\right)\right|\\
 & = & \frac{1}{2}\ln\left[\frac{\cosh^{2}\left(\frac{\pi}{2a}(y-b)\right)-\cos^{2}\left(\frac{\pi x}{2a}\right)}{\cosh^{2}\left(\frac{\pi}{2a}(y-b)\right)-\sin^{2}\left(\frac{\pi x}{2a}\right)}\right].
\end{eqnarray}
Therefore, when considering two lines of alternate disclinations as in Fig. \ref{row}, $V(x,y)$  becomes: 
\begin{eqnarray}
V(x,y)=\frac{|F|}{4\pi}\ln\left[\left(\frac{\cosh^{2}\left(\frac{\pi}{2a}(y-b)\right)-\cos^{2}\left(\frac{\pi x}{2a}\right)}{\cosh^{2}\left(\frac{\pi}{2a}(y-b)\right)-\sin^{2}\left(\frac{\pi x}{2a}\right)}\right)\left(\frac{\cosh^{2}\left(\frac{\pi}{2a}(y+b)\right)-\sin^{2}\left(\frac{\pi x}{2a}\right)}{\cosh^{2}\left(\frac{\pi}{2a}(y+b)\right)-\cos^{2}\left(\frac{\pi x}{2a}\right)}\right)\right].\ \label{fhat}
\end{eqnarray}
We can think of this function as a sort of gravitational ``potential" acting on point masses that move in the presence of the arrays of defects (see, for example, reference \cite{grats}). Eq. (\ref{fhat}) together with (\ref{V}) in (\ref{metric}) specifies the metric of the effective geometry associated to the array of disclinations. Even though single defect metrics are well known in the literature \cite{Puntingam}, to the best of our knowledge, this is the first time a metric for a long array of disclinations  is given. We remark that this particular array was studied with very different purposes in ref. \cite{Cordier14}.

\begin{figure}[h!]
\centering \includegraphics[height=9cm]{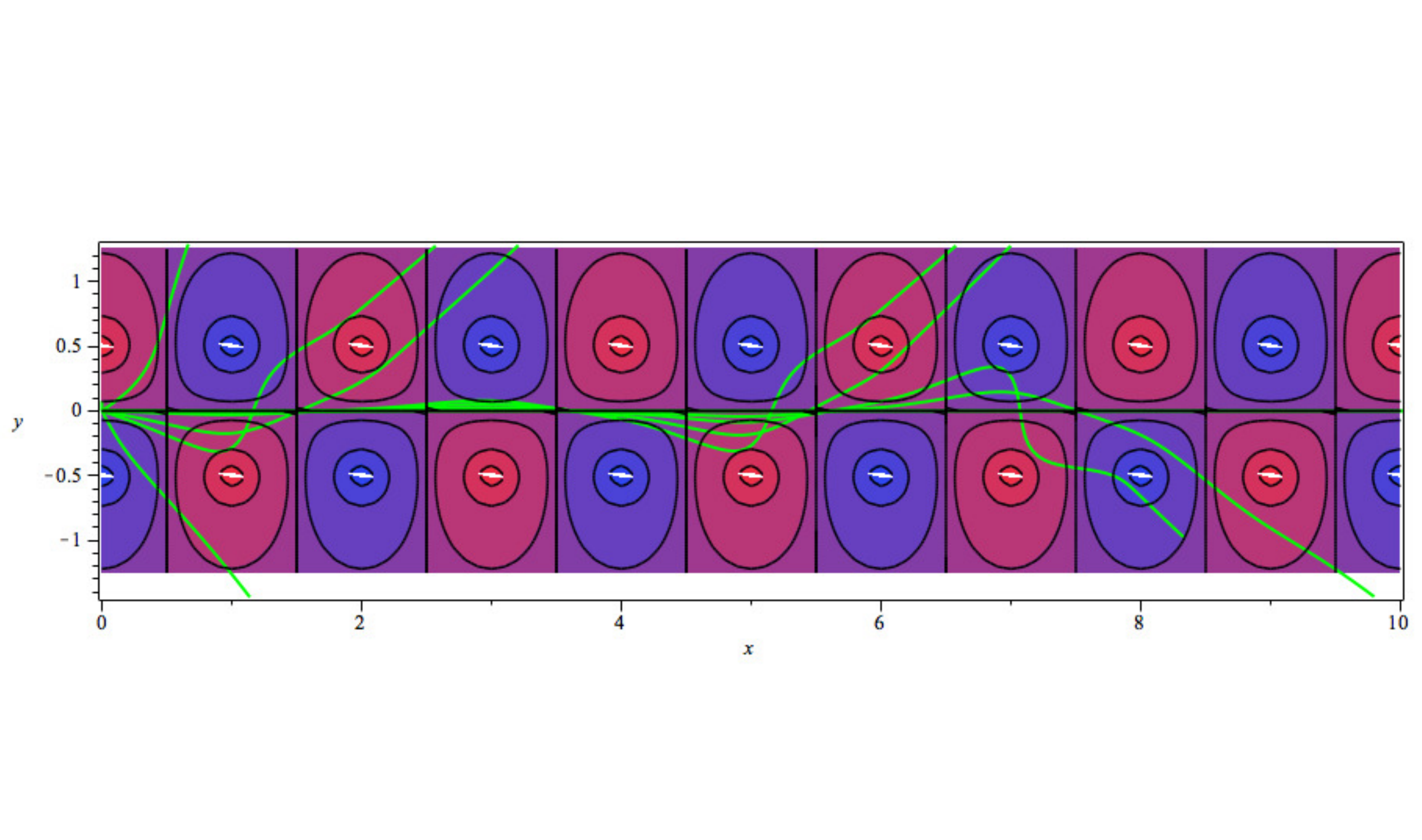} \vspace{-1cm}
 \caption{Different geodesics, shot from the origin, in the channel of disclinations geometry. The positive disclinations correspond to red contours while negative disclinations correspond to blue contours. Depending on the shooting angle, the propagation
of phonons may be guided by the street of topological defects.}
\label{FigGeodesics1} 
\end{figure}

\begin{figure}[h!]
\centering \includegraphics[width=1\linewidth]{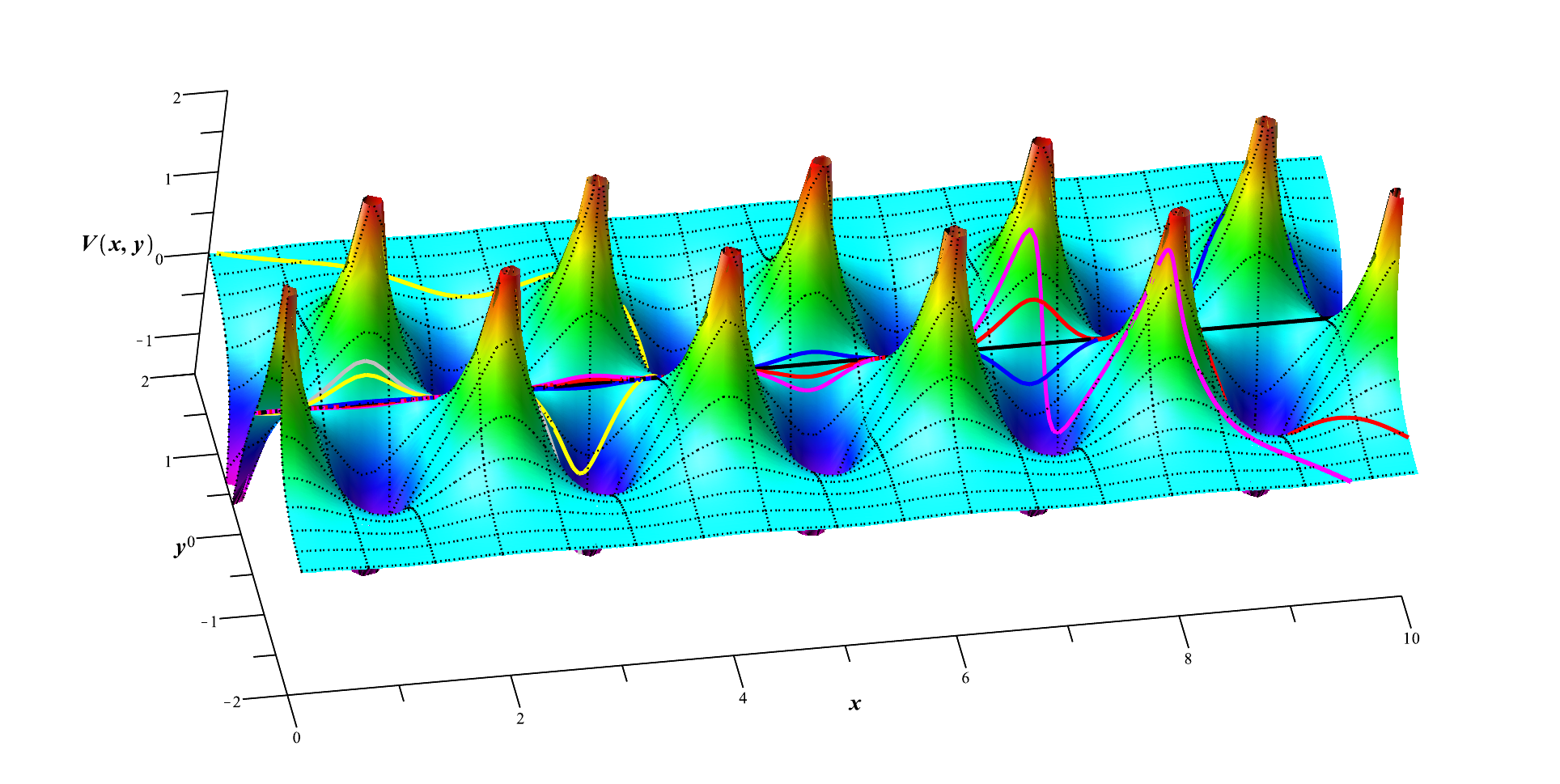} 
\caption{Geodesics of the channel of defects in the potential landscape. }
\label{Fig3dGeodesics} 
\end{figure}

In the small wavelength approximation (or geometrical optics limit), the paths followed by phonons are no longer straight lines, but they are the geodesics of the channel of disclinations geometry, that is the trajectories of shortest lengths in the background geometry. Indeed, according to \cite{Katanaev05}, deviation of the effective geometry to Euclidean space stems from the strain tensor (or equivalently, the stress tensor, from Hooke's law) and therefore, phonons can be understood as free-falling particles in the background geometry. Hence, they obey the so-called geodesic equations:
\begin{equation}
\frac{d^{2}x^{\mu}}{d\lambda^{2}}+\Gamma_{\rho\sigma}^{\mu}\frac{dx^{\rho}}{d\lambda}\frac{dx^{\sigma}}{d\lambda}=0,\label{geodesic}
\end{equation}
where $\lambda$ is an affine parameter along the path and $\Gamma_{\rho\sigma}^{\mu}$ are the coefficients of the  Levi-Civita connection which can be expressed from the metric tensor components as 
\begin{equation}
\Gamma_{\rho\sigma}^{\mu}=\frac{g^{\mu\nu}}{2}\left(\partial_{\rho}g_{\nu\sigma}+\partial_{\sigma}g_{\nu\rho}-\partial_{\nu}g_{\rho\sigma}\right)\label{christoLC}
\end{equation}
since there is no torsion. After some calculations, (\ref{geodesic}) reduce to \cite{Letelier01}: 
\begin{eqnarray}
 &  & \ddot{x}-2\left(\dot{x}^{2}-\dot{y}^{2}\right)\partial_{x}V-4\dot{x}\dot{y}\partial_{y}V=0, \nonumber \\
 &  & \ddot{y}-2\left(\dot{x}^{2}-\dot{y}^{2}\right)\partial_{y}V-4\dot{x}\dot{y}\partial_{x}V=0.\label{geodesic2}
\end{eqnarray}
Although the geodesics' system of differential equations above are analytical by nature, the inherent complexity of $f_{V}(x,y)$ in Eq. (\ref{eqf}) leads us to cumbersome expressions
for the system in Eq. (\ref{geodesic2}). We circumvent such complexity by using symbolic algebra, thus yielding a quite complex system of differential equations for the geodesics, as illustrated in a flexible Maple code, available from the authors upon request, for arbitrary $F$, $a$, and $b$.  In order to get both quantitative and qualitative informations about the geodesics in the channel, we systematically solved the system of geodesic equations. We fixed $F=\pm \pi$, $a=1,b=0.5$  (arbitrary distance units) and solved numerically the resulting system of differential equations. This choice of parameters was made by convenience, in order to emphasize the effects of the defect array as shown in the graphs. It was not inspired in any particular material.
Without loss of generality, we chose geodesics starting at the origin, $(x_{0}(0)=0,y_{0}(0)=0)$, with unitary initial ``speed", $x'(0)=\cos(\theta_{0}),y'(0)=\sin(\theta_{0})$,
which defines the shooting angle. In Fig. \ref{FigGeodesics1} we illustrate some geodesics in the channel of defects for a few shooting angles. In Fig. \ref{Fig3dGeodesics} we depict some geodesics in the potential landscape. 

\noindent 

The phononic paths, as represented by the geodesics seen in Figs. \ref{FigGeodesics1} and \ref{Fig3dGeodesics}, are very sensitive to the shooting angle. Furthermore, they appear to be attracted by the positive defects while repelled by the negative ones. This is quite natural, since the former have the geometry of a cone (with less space than the plane) and the latter of an anticone (with more space than the plane). This makes the geodesic bend toward the positive defect and away from the negative defect. In other words, the rays give a picture of how phonon propagation occurs in the medium with an array of defects. We can infer that optimized distribution of defects may be used to design devices like thermal lenses or heat waveguides, as phonon flux undergo total internal reflections when approaching disclinations \cite{Fumeron08}.

\section{Phonon Boltzmann transport equation and Clausius invariant}

\begin{figure}[h!]
\centering \includegraphics[height=2in]{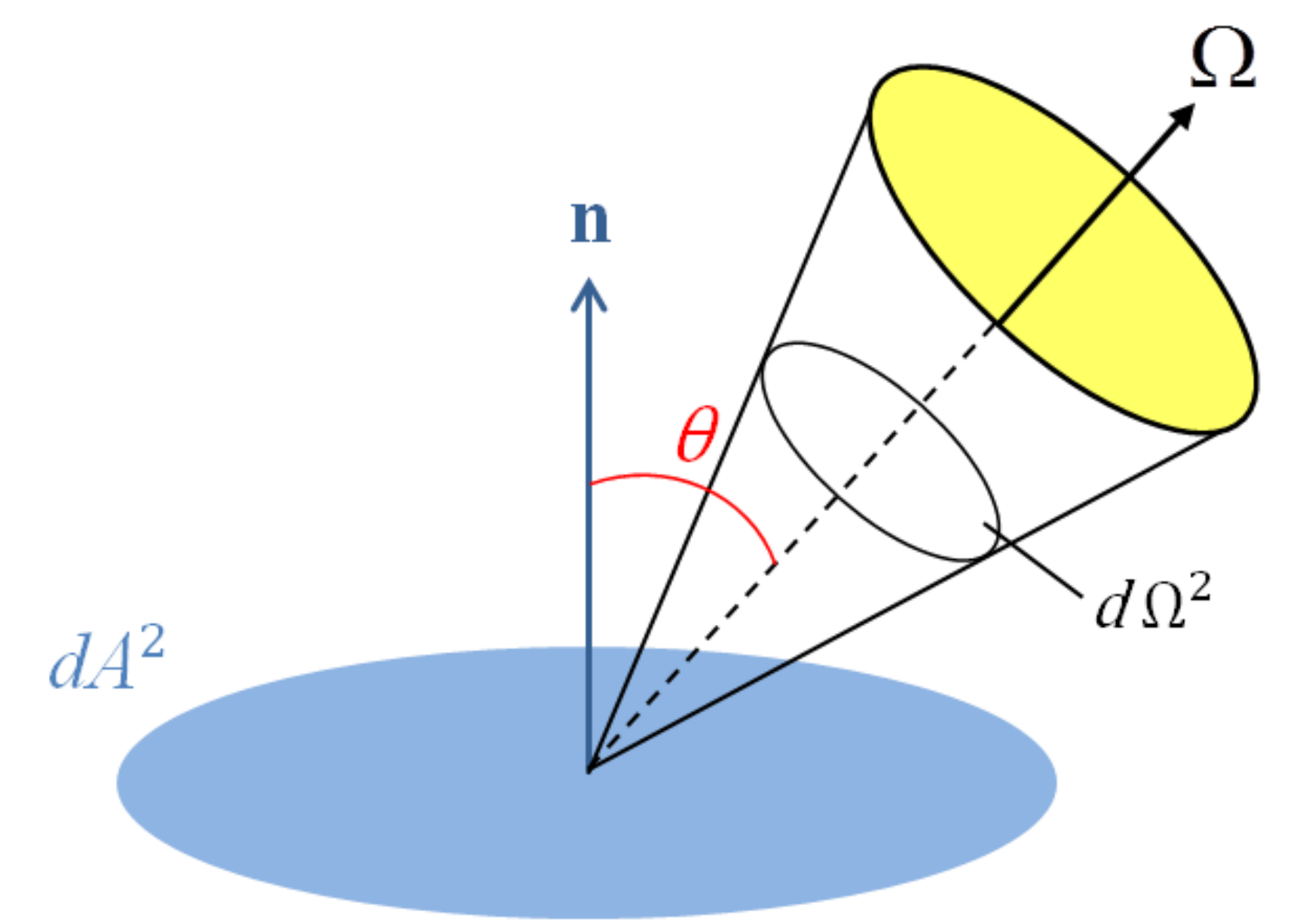} \caption{Definition of the specific intensity.}
\label{rad} 
\end{figure}

Transport of phonons is governed by the BTE, which shares common features with older radiative heat transfer theories. Originally, radiative transfer consists in a phenomenological description
for the interactions of photons with participating matter. Based on the pioneering works by Khvolson \cite{Khvolson90}, Schuster \cite{Schuster05} and Schwarzschild \cite{Schwarzschild06},
this theory borrows concepts from radiometry (measurable quantity such as the radiative flux), classical optics (Fermat principle to get the optical paths) and quantum theory (Planck blackbody function
to account for the thermal emission by matter). The central quantity is the specific intensity $I_{\nu}$ that corresponds to the energy conveyed by an amount $dN$ of carriers crossing an element $dA$ of area during time $dt$, in a frequency range between $\nu$ and $\nu+d\nu$, and within an element $d\Omega$ of unit solid angle centered on direction $\boldsymbol{\Omega}$ according to \cite{Modest03} (see Fig. \ref{rad}):
\begin{equation}
I_{\nu}=\frac{h\nu dN}{\cos\theta\: dt\: dA\: d\nu\: d\Omega} \label{Iradio}
\end{equation}

The equation governing $I_{\nu}$ is derived from a simple energy balance along any energy path: specific intensity decays ought to absorption and scattering processes, and
it gets stronger because of scattering and inner source terms. In its usual form, it is therefore given by \cite{Modest03,Chandra60}:
\begin{equation}
\frac{dI_{\nu}}{dl}=\frac{I^{0}\left(T\right)-I_{\nu}}{\Lambda}, \label{RTE}
\end{equation}
where $d/dl$ is the curvilinear arc length, $I^{0}$ is the equilibrium intensity, $T$ is the temperature and $\Lambda$ is the phonon mean free path in the non-defected solid.

Despite its handiness, the phenomenological approach hides some intricate points. Indeed, anytime a symmetry of physical properties of the participating medium is lost, (\ref{RTE}) must be modified. As a matter of fact, in nonstationnary and graded-index media, the adiabatic invariant along a path is no longer the specific intensity but the so-called Clausius invariant \cite{Kravtsov96}: 
\begin{equation}
C_{\nu}=\frac{I_{\nu}}{n(\mathbf{r})^{2}\nu^{3}},\label{clausius}
\end{equation}
where $n\left(\mathbb{r}\right)$ is the refractive index (the same result was also found from other arguments by \cite{Marechal68,MTW73,Pomraning73,Fumeron08}). It must be emphasized that (\ref{clausius}) also holds for radiative transfer of elastic waves, as an analog of Fermat principle can be obtained in elastodynamics \cite{Horz95}. Hence, considering the balance on Clausius invariant, the modified BTE now writes as:
\begin{equation}
\frac{dC_{\nu}}{dl}+\frac{C_{\nu}}{\Lambda}=\frac{I_{\nu}^{0}\left(T\right)}{\Lambda}. \label{MRTE}
\end{equation}
It is important to recall that in (\ref{MRTE})), the balance is performed along ray paths. Whether it is beams of photons or of phonons \cite{Horz95}, the ray paths obey Fermat principle which, as previously stated, requires to determine the refractive index. In the next section, we introduce the distribution of disclination dipoles from a geometric standpoint and extract from it the effective refractive index experienced by phonons.

\section{Boltzmann transport equation in the channel of disclinations geometry}

Based on the two preceding sections, we are now focusing
on the modified BTE in the presence of the array of disclinations
and determine some of its most remarkable properties. First, we briefly
present the general procedure to obtain the effective refractive index
from a diagonal metric. Consider a wave propagating along direction
$\boldsymbol{\Omega}=(\sin\hat{\theta}\cos\hat{\phi},\sin\hat{\theta}\sin\hat{\phi},\cos\hat{\theta})$
($\hat{\theta}$ and $\hat{\phi}$ denote the polar and azimuthal
angles) in a geometric background described by the metric tensor $g_{\mu\nu}$,
then the dispersion relation of a wave associated to 4-wavevector
$K^{\mu}=(\omega/c,k\sin\hat{\theta}\cos\hat{\phi},k\sin\hat{\theta}\sin\hat{\phi},k\cos\hat{\theta})$
writes as $K^{\mu}g_{\mu\nu}K^{\nu}=0$, which gives 
\begin{equation}
-\frac{\omega^{2}}{c^{2}}\left|g_{00}\right|+k^{2}\left(g_{11}\sin^{2}\hat{\theta}\:\cos^{2}\hat{\phi}+g_{22}\sin^{2}\hat{\theta}\:\sin^{2}\hat{\phi}+g_{33}\cos^{2}\hat{\theta}\right)=0,\label{a1}
\end{equation}
(all space components $g_{ii}$ are positive). On the other hand,
the dispersion relation of a wave in a dielectric medium at rest and
of refractive index $n$ is given by: 
\begin{equation}
-\frac{\omega^{2}}{c^{2}}n^{2}+k^{2}=0.\label{a2}
\end{equation}
Substituting $k^{2}$ from (\ref{a2}) into (\ref{a1}) defines an
effective refractive index 
\begin{equation}
n\left(\bold{r},\boldsymbol{\Omega}\right)=\sqrt{\frac{\left|g_{00}\right|}{g_{11}\sin^{2}\hat{\theta}\:\cos^{2}\hat{\phi}+g_{22}\sin^{2}\hat{\theta}\:\sin^{2}\hat{\phi}+g_{33}\cos^{2}\hat{\theta}}}.\label{a3}
\end{equation}
Generally speaking, this refractive index depends both on the position
in the medium through the radial vector $\mathbb{r}$ and the local
propagation direction $\boldsymbol{\Omega}$, but it does not exhibit
any dispersion.

\begin{figure}
\centering \includegraphics[width=1\linewidth]{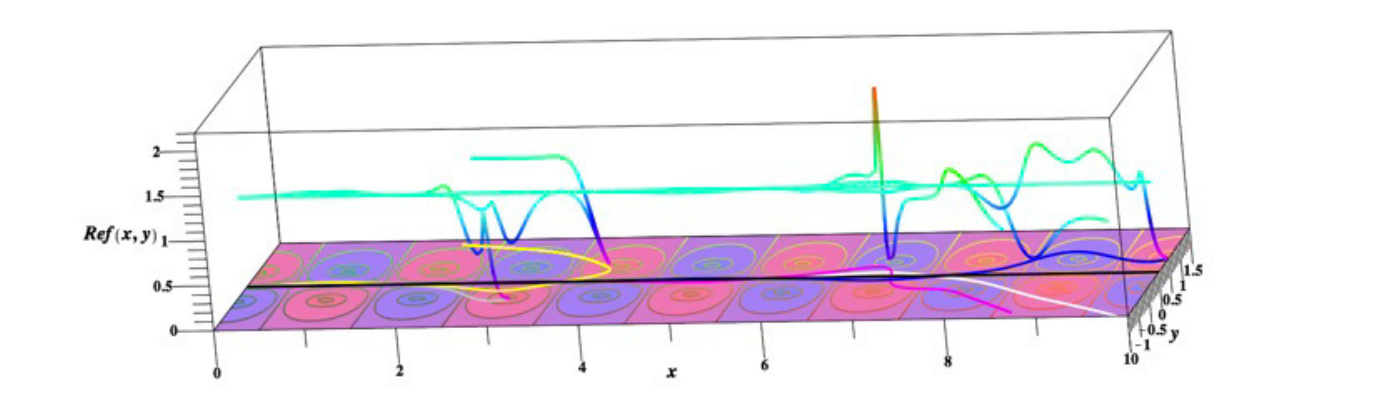} \vspace{-1cm}
 \caption{Refraction index projected on channel of disclinations along some
representative geodesics. The refraction index is nearly constant
far from the defects and exhibits strong gradients close to them.}
\label{FigGeodesics} 
\end{figure}

In the case of the channel of disclinations, the effective refractive
index writes as 
\begin{equation}
n(\bold{r},\hat{\theta})=\frac{1}{\sqrt{e^{-4V(x,y)}\sin^{2}\hat{\theta}+\cos^{2}\hat{\theta}}}.\label{effn}
\end{equation}
Notice that the refractive index depends both on the position $(x,y)$
and on the local propagation direction $\hat{\theta}$. In Fig. \ref{FigGeodesics}
a plot of $n(\bold{r},\hat{\theta})$ along selected
geodesics is presented.  A thorough examination of the light paths shows that in between defects, the refractive index is constant and rays propagate along straight lines, whereas close to defects, light trajectories present strong curvature. As expected from Fermat principle, the bending occurs in such a way that the center of curvature belongs to the region of higher refractive index.

While the metric exhibits translational symmetry along the $z$ axis,
hence the coefficients $g_{\mu\nu}$ do not depend on $z$, this is
not true for the effective refractive index which looses this symmetry
through the dependence on the wavevector direction. From this perspective,
Eq.(\ref{effn}) might be misleading since only the angle $\hat{\theta}$
appears explicitly, but the presence of the azimuthal angle $\hat{\phi}$
is implicit in the $x,y$ dependence. From (\ref{clausius}), one
deduces the corresponding Clausius invariant 
\begin{equation}
C_{\nu}\left(t,\mathbf{r},\boldsymbol{\Omega}\right)=\frac{I_{\nu}\left(t,\mathbf{r},\boldsymbol{\Omega}\right)}{\nu^{3}}\left(e^{-4V(x,y)}\sin^{2}\hat{\theta}+\cos^{2}\hat{\theta}\right).
\end{equation}
Therefore, the modified BTE (MBTE) in the presence of the array of
defects is given by 
\begin{eqnarray}
\frac{d}{dl}\left[\frac{I_{\nu}\left(t,\mathbf{r},\mathbf{\Omega}\right)}{\nu^{3}}\left(e^{-4V(x,y)}\sin^{2}\hat{\theta}+\cos^{2}\hat{\theta}\right)\right]+\frac{1}{\Lambda}\frac{I_{\nu}\left(t,\mathbf{r},\mathbf{\Omega}\right)}{\nu^{3}}\left(e^{-4V(x,y)}\sin^{2}\hat{\theta}+\cos^{2}\hat{\theta}\right)=\frac{I_{\nu}^{0}\left(T\right)}{\Lambda}.\label{MRTE-a}
\end{eqnarray}

Some physical insights can be obtained by comparing Eq. (\ref{MRTE-a}) to the one without defects (\ref{RTE}). First, let us consider the simple steady-state case in a cold medium which writes phenomenologically as 
\begin{eqnarray}
\frac{dI_{\nu}}{dl}+\left[\frac{1}{\Lambda}+\left(\frac{d}{dl}\ln\: \left[e^{-4V(x,y)}\sin^{2}\hat{\theta}+\cos^{2}\hat{\theta}\right]\right)\right]I_{\nu}=0, \label{psBL}
\end{eqnarray}
or more simply
\begin{eqnarray}
\frac{dI_{\nu}}{dl}+\left[\frac{1}{\Lambda}+\frac{1}{\mathcal{L}_d}\right]I_{\nu}=0. \label{psBL}
\end{eqnarray}
Besides the bulk mean free path $\Lambda$, a new length scale for the scattering of phonons, $\mathcal{L}_d$, arises as consequence of the channel of defects. It is given by 
\begin{eqnarray}
\frac{1}{\mathcal{L}_d\left(x,y,\hat{\theta}\right)} & = & \frac{2\sin\hat{\theta}}{\left(e^{-4V(x,y)}-1\right)\sin^{2}\hat{\theta}+1}\left[2\sin\hat{\theta}\: e^{-4V(x,y)}\left(\frac{\partial V}{\partial x}\frac{dx}{ds}+\frac{\partial V}{\partial y}\frac{dy}{ds}\right)\right.\nonumber \\
 &  & \left.\qquad+\cos\hat{\theta}\left(e^{-4V(x,y)}-1\right)\frac{d\hat{\theta}}{ds}\right].\label{inh}
\end{eqnarray}
The first derivatives $dx/ds$, $dy/ds$ and $d\hat{\theta}/ds$ are obtained numerically from the geodesic equations (\ref{geodesic}). There is no constraint to settle the signs of the different terms
involved in $l_{\alpha}$: hence, this parameter can be either of positive or of negative sign, so that the array of defects either damps or amplifies locally the specific intensity. That means that
the distribution of defects does not only curve the geodesics, but it also leads to a local geometric reduction/enhancement of energy carried by the waves. Although (\ref{psBL}) is analog to a Beer-Lambert law, it must be emphasized that there is no absorption process here. The scattering length due to the disclinations (\ref{inh}) rather involves the local spatial enhancement (where it is negative) or local spatial reduction (where it is positive) of the energy carried by the phonons, like a focusing/defocusing effect. Curvature doping of radiant intensity is well-known in radiative transfer
and was first discussed in \cite{PBA02}. 

\section{Concluding remarks}

To sum up, we proposed a simple toy-model for phonon transport through a distribution of disclinations. In the presence of two rows of alternate wedge disclinations, two new phenomena arise. First, phonon paths are geometrically scattered by an effective background geometry, presenting alternate regions of positive and negative curvature originating from the alternate disclinations themselves. Second, as testified by the change in Clausius invariant, the energy carried by beams can be geometrically amplified/attenuated depending on the direction of the propagation. As these two effects are competing against each other, the heat transfer problem is expected to be extremely sensitive to the kind of thermal constraints set on the sample (Neumann condition or Dirichlet conditions) and to the channel parameters as well (extreme sensitivity to shooting conditions is manifest from Fig. 4).

The main objective of this work is to show that, with a geometric approach such as the one  presented here, one can study phonon transport in media with arrays of topological defects. The effective geometry gives not only the ray paths but also the refractive index as function of position and direction of propagation, a result of the anisotropy introduced by the defects.  This opens up possibilities of designing phonon transport devices (thermal devices) that may include effective phonon manipulation  by defect engineering.  In the specific example studied here,  a street of alternate disclinations, appears as a channel for phonon propagation with strong influence on the refractive index, causing local focusing/defocusing of the beam. Moreover, an asset of this formalism is that although polarization states of the waves and mode coupling effects were not explicitly discussed, they can formally be incorporated
from the above results to obtain the generalization of the BTE. Investigating their impact on transfer will be the object of our further investigations.

\noindent
\textbf{Acknowledgements}: FM, FS and EP are indebted to CNPq, CAPES and FACEPE (Brazilian agencies) for their financial support. FM also thanks Universit\'e de Lorraine for a Guest Lecturer award. We also thank E. Gomes for helping with figure 2.

\end{document}